\documentclass[twocolumn,showpacs,preprintnumbers,amsmath,amssymb,9pt]{revtex4}

\usepackage{graphicx}
\usepackage{dcolumn}
\usepackage{bm}

\begin{document}
\title{Probing Low-x QCD With Very High Energy Prompt Muons}
\author{Namit Mahajan$^{a\ast}$ and Sukanta Panda$^{b\dag}$}
 \vskip 1cm
\affiliation{{$^a$\em Department of Physics,
         National Taiwan University, Taipei 10617, Taiwan (ROC).}\\
{$^b$\em Departamento de Fisica Teorica C-XI and Instituto de Fisica
Teorica C-XVI,}\\
{\em  Universidad Autonoma de Madrid, Cantoblanco, E-28049 Madrid,
Spain.}\\
{{\em E-mail: }$^{\ast}$nmahajan@phys.ntu.edu.tw,
$^{\dag}$sukanta@delta.ft.uam.es}}

\textheight 10in \textwidth 7in

\begin{abstract}
We explore the possibility of utilizing the prompt muon fluxes at
very high energies in order to discriminate various
models/parametrizations of low-x QCD behaviour of hadronic
cross-sections relevant at such energies. We find that the pair
meter technique for measuring high energy prompt muons can be very
efficient in such an endeavor. As a by product, it allows to cleanly
probe the change in composition of the primary cosmic rays expected
at high energies.
\end{abstract}



\maketitle

{\bf Introduction~~} The cosmic ray (CR) spectrum is characterized
by a sharply falling power law behaviour, $\frac{dN}{dE} \sim
E^{-(\gamma + 1)} $\cite{gai}. The spectrum gets more steeper around
$10^6$ GeV with the spectral index $\gamma$ changing from $1.7$ to
$2.1$ - this region is called the {\it knee}. Around $E\sim 5\times
10^9$ GeV, one observes a flattening of the spectrum, with the
spectral index $\gamma$ falling between $1.4$ and $1.7$. This is the
so called {\it ankle}. The change in the slope of the spectrum at
these two places is a puzzling issue. The region beyond the ankle is
the regime of ultra high energy cosmic rays. There is not much data
available in that region and no clear consensus exists on the
composition or the particle content in this region \cite{nag}. It is
generally believed that the change in the slope around the knee is
astrophysical in nature rather than any specific change in hadronic
properties and/or interactions. Though not conclusive, there appears
to be some evidence \cite{kas3} that the composition is heavier
above the knee region. If true, then a
 significant suppression of the very high
energy ($\geq 10^5$ GeV) lepton fluxes is expected due to CR
interactions during the journey downwards\cite{can}.

There is a sharp reduction of the lepton flux from pions and kaons
(this component is called the {\it conventional flux}) above a few
TeV \cite{lip}. This is due to the increasing competition between
their interaction and decay lengths. Therefore, at very high
energies, almost all the lepton flux is expected to arise from the
semi-leptonic decays of heavy hadrons (this component is called the
{\it prompt flux}), most noticeably the charmed hadrons. It must be
mentioned that the B-hadrons can also contribute significantly to
the lepton fluxes in this energy range. This perhaps becomes an
important issue for the $\nu_{\tau}$ because the only source of
$\nu_{\tau}$ in the charm sector is $D_s$ while there can be many
B-hadron decay channels giving rise to $\tau$ and $\nu_{\tau}$.
Another point of importance is the fact that the neutrinos
propagating at such high energies have almost vanishing
probabilities to oscillate to another flavour. Therefore,
measurement of neutrino flux in this regime requires no corrections
due to the oscillation phenomenon. Further, it is known that the
prompt muon flux is only about $10\%$ smaller than the prompt
$\nu_{\mu}$ flux at the surface of the earth. Therefore, measurement
of the prompt muon flux at high energies will act as a normalization
for the prompt neutrino flux and a direct comparison of the two is
both desirable and also necessary. The prompt muon flux, thus, is an
object of great importance, not just from the above mentioned points
but also from the fact that such muons are a major background to
various neutrino experiments.

To this end, we require precise theoretical predictions for the
prompt lepton fluxes. However, in reality the situation is
drastically the opposite. Various phenomenological predictions for
the prompt lepton fluxes span over two decades of interval
\cite{tig,zhv,rvs,prs,bnsz,ggv2, mrs}. The reason is mainly due to
the vastly different choices for the charm production cross-section
- perturbative QCD (pQCD) with a $K$ factor \cite{tig},
next-to-leading order (NLO) pQCD \cite{ggv2,prs}, quark-gluon string
models and recombination quark-parton models \cite{bnsz}. Within the
QCD based models, the problem can be traced back to the fact that
the available data from accelerator/collider experiments is used as
an input for the charm production (and also for other hadrons) at
those experiments. This is then extrapolated to high energies and
low-x values. Typical values sampled in high energy cosmic ray
events, relevant for prompt muon flux via the charm production,
correspond to $x \sim 10^{-5}$ and maybe smaller. There is no data
available in this region and one is forced to extrapolate without
clear guidelines. The gluon distribution function $g(x)$ extracted
from the available data, and naively extrapolated to higher energies
and lower x values, is known to grow indefinitely, thereby causing
concern regarding unitarity. Physically, it is expected that some
mechanism should tame this sharply rising behaviour, though none is
conclusively known at present. Often, the theoretical models assume
a power law behaviour for the gluon distribution function, $ xg(x)
\sim x^{-\lambda},$ with $\lambda$ varying in the range $0-0.5$. The
fluxes therefore depend strongly on  the chosen value of $\lambda$.
The data from $ep$ scattering at HERA, albeit at a very different
energy and low-x regime, show a rise in the gluon distribution
function. On the other hand, RHIC data at higher energies and for
different compositions of the colliding particles (also later HERA
data) gives some indication of partial taming of the rising
distribution function. This is the so called {\it saturation} of
gluon density at very low-x. Roughly speaking, at very low-x values,
one can expect that the non-linear effects, recombination effects to
become significant, thereby leading to the saturation or taming of
the gluon density. For an early discussion of RHIC results and
theoretical expectations, see \cite{rhic1}. The saturation models
have been shown to fit the HERA data as well. An important lesson is
to be very cautious in extrapolating the existing accelerator data
to cosmic ray energies. The saturation picture can be expected to be
applicable at cosmic ray energies\cite{dds}. For a general overview
of various theoretical issues in low-x and saturation physics see
\cite{lowx1}. An overview of hadronic interactions and cosmic rays
can be found in \cite{lowxcr}.

The experimental situation is not very precise either at this stage.
Various experiments {\cite{lvd}} provide upper limits on the muon
and neutrino fluxes in the energy range of interest. At present,
these limits allow a large variation in the prompt fluxes. One can
therefore expect that better measurements of muon fluxes can play a
definitive role in selecting the charm production models, and
thereby, also providing invaluable information about parton
densities at such low-x and high energy values. Another related
source of large theoretical uncertainties is strong dependence of
the hadronic cross-sections on the renormalization and factorization
scales. This is partly related to the naive extrapolation of parton
distribution functions to very different energy and x-values. For
the case of conventional fluxes originating from the pions and
kaons, these issues are in much better control and therefore the
predictions stand on a sound footing.

In the present letter, we explore the possibility of utilizing the
high energy prompt muon flux(es) in order to investigate whether the
general expectations expressed above can in practice help in
selecting the charm production model/parametrization. The range of
models or parameterizations employed in the literature is too vast
to be covered in this work. We choose some of the models often used
and compare the predictions, incorporating the saturation model of
Golec-Biernat and Wuthsoff \cite{gbw}. Also, it should be mentioned
that for the case of muons and $\nu_{\mu}$, the flux from the beauty
hadrons is not more than a few percent of that from the charm
hadrons. Therefore, we ignore the beauty contributions in this work.
However, for $\nu_{\tau}$ flux, there is almost $40\%$ enhancement
\cite{mrs} and should be considered.

{\bf Prompt muons and their detection~~} At higher and higher
energies, the (prompt) muon flux is expected to reflect the onset of
contributions from the production of heavy hadrons - the charmed and
the beauty. The energy dependence of prompt muons (the flux is
isotropic) follows the parent cosmic ray spectrum and is therefore
harder. In contrast, the conventional  flux energy spectrum is more
steeply falling (almost by one extra power), caused due to the
tension between the decay and interaction of the parent mesons while
passing through the atmosphere. A schematic picture of the entire
process, very similar to the conventional lepton flux, is as
follows: \\
$CR ~flux (X=0)\to Flux (X) \to Charm ~production \to $\\ $\to Decay
\to Prompt ~Flux \to To ~detector$ \\Of all the ingredients entering
the calculation, the model for charm production has the biggest
uncertainties as discussed above. Before proceeding further, let us
have a look at the expression for the lepton flux. Assuming infinite
isothermal atmospheric depth, the final expression for the flux, for
lepton energies $E_l< \epsilon_{charm}^{crit}\approx 10^7$ GeV is
(for zenith angle $< 60^{\circ}$, $\epsilon_i^{crit} =
\frac{m_ih_0}{c\tau_i}$, $h_0=6.4$ km where $m_i,\tau_i$ are the
rest mass and life time of the $i$th particle)
\begin{equation}
\phi_l(E_l) = \underbrace{Z_{Y_cl}(E)}_{Y_c\to
l}\underbrace{Z_{NY_c}(E)}_{N+A\to Y_c}
\underbrace{\frac{\Lambda_N(E)}{\lambda_N (E)}}_{depletion} \phi_N
(E,0)
\end{equation}
 where $\Lambda_N(E)= \frac{\lambda_N(E)}{1-Z_{NN}(E)}$ and
 $\lambda_N(E)=\frac{\rho_{atm}(X=h)}{\sigma_{NA}n_{A}(X=h)}$
  are the attenuation and interaction
length of the nucleon in the atmosphere. $\phi_N(E,0)$ is flux of
the nucleon at depth X=0. $Y_c$ denotes various charm hadrons in the
sequence - $D^{0(\pm)}, D_s, \Lambda_c$. For the total Nucleon-Air
cross section, we adopt the parametrization \cite{miel}
\begin{equation}
\sigma_{NA}(E)= \left[280 - 8.7 \ln\left( \frac{E}{GeV}\right) +
1.14 \ln^2\left( \frac{E}{GeV}\right)\right] mb
\end{equation}
Assuming scaling, spectrum weighted
moments for a power law initial flux, $\phi_N(E,X=0) = \phi_{0N}
E_N^{-(\gamma + 1)}$, are defined as
\begin{equation}
Z_{Nj}(E) = \int_0^1 dx x^{\gamma -1}
\frac{1}{\sigma_{NA}(E)}\frac{d\sigma_{NA \rightarrow j}(E,x)}{dx}
\end{equation}
where $x\approx \frac{E_j}{E}$ and $j$ denotes any of the final
species. The charm hadron Z moments are related to the partonic
level charm Z moments through their relative yields \cite{mrs,fmnr}, $Z_{Nj}(E) = f_{j} Z_{Nc}(E)$.
The lepton production Z moments can be defined in an analogous
fashion by replacing the production cross-section by the
(differential) decay rates and plugging in the corresponding
branching fractions \cite{sine1}.

It is therefore quite clear from the above expressions that the
lepton fluxes at the end are quite sensitive to the charm production
cross section. Till the knee, the cosmic ray flux and composition is
rather established and therefore, the only source of large error is
the charm cross section. This therefore gives us a unique
possibility to gain information about heavy quark production
mechanism at high energies and low-x. Due to very penetrating nature
of the muons, it is extremely hard to have a direct detection. The
direct measurement of such high energy muons will require
impractically large detectors with, presumably, very strong magnetic
fields. Also the error in magnetic deflection of the muons at such
large energies become comparable to the total angular deflection
itself. However, there exists a clean method capable of capturing
the energy information of the incoming high energy muons. The
method, {\it pair meter technique} {\cite{alek}}, is based on
measuring the individual muon energy via the energy deposition by
fast muons in the dense matter. It is quite well known that above a
few TeVs, the dominant energy loss mechanism for muons while passing
through matter is the $e^+e^-$ pair production, valid for the whole
range of momentum transfer values. Further, at TeV energy or above,
the differential energy loss, $dE_{\mu}/dx$ is linear in $E_{\mu}$,
providing a clean estimation of muon energy. This technique has been
successfully employed for a limited cosmic ray data \cite{ccfr} and
the study confirms that the same method is capable of being used for
even high energy muons. The requirement for such a measurement is to
have the muons pass through a dense material like iron (or some
other heavier material). In this regard, it is worthwhile to
emphasize that the typical expressions employed in most cosmic ray
studies are Born cross sections. However, for heavier materials,
there is a need to consider the Coulomb corrections arising due to
all order resummation effects coming from $Z\alpha_{em}$, where $Z$
is the atomic number of the material \cite{ivanov}. One can easily
verify that for the case of iron, even at such large muon energies
and for small momentum transfer, these corrections are at most a
couple of percent. For lead, on the contrary, the Coulomb
corrections are way too large and can not be neglected. Therefore,
iron seems to be ideal for such a case. We, thus, choose to
investigate the capability of this method in context of an iron
calorimeter. For the sake of illustration we choose to work with a
volume of $50$ kT. This choice is not completely arbitrary but
motivates from an iron calorimeter detector planned in India
\cite{ino}. However, we would also like to emphasize that the
features of the results and conclusions obtained below are going to
remain the same for some other choice as well. Furthermore, in the
study below, we do not commit to very specific topology of the
detector and the surroundings. Therefore, one can expect some
calculable changes in the results for specific details of the
detector considered.

It is noteworthy that the underwater or under ice experiments like
BAIKAL\cite{baikal} and AMANDA are capable of, and already measuring the
zenith
angle and depth dependence of the (integrated) muon flux. This is to
be contrasted with the proposal to employ pair meter technique to
measure the energy spectrum of muons, with the latter giving direct
information about energy of individual muons. Measurement of muon
energy spectrum will be of immense use in directly addressing issues
related to cosmic rays and also high energy neutrino physics. We
believe that a combination of results from both type of methods will
play a decisive role in settling various issues.


{\bf Muon fluxes in different models of charm production~~} As we
discuss in the introduction there exist a host of studies on the
prompt muon fluxes Since it is not possible to provide an exhaustive
comparison in this letter, we choose to survey some of the cases.

Following {\cite{tig}} the conventional flux (CONV) and the prompt
flux (TIG) have been parameterized as
\begin{eqnarray}
\frac{dN}{dE} = \left\{\begin{array}{ll} \frac{N_0 E^{-\gamma -
1}}{1 + A E}& \textrm{$E < E_{a}$}\\
 \frac{N_0^{'} E^{-\gamma^{'} - 1}}{1 + A E}
& \textrm{$E > E_{a}$} \end{array}\right.
\end{eqnarray}
For the conventional muon flux $N_0 = 0.2,$ $N_0^{'} = 0.2,$
$\gamma=1.74, $ $\gamma^{'} = 2.1,$ $E_a = 5.3 \times 10^5$ GeV, $A
= 0.007$ while for the prompt muon flux $N_0 = 1.4 \times 10^{-5},$
$N_0^{'} = 4.3 \times 10^{-4},$ $\gamma=1.77, $ $\gamma^{'} = 2.01,$
$E_a = 9.2 \times 10^5$ GeV, $A = 2.8 \times 10^{-8}.$

Gelmini, Gondolo and Varieschi (GGV)\cite{ggv2} have included NLO
corrections for the charm production with $ xg(x) \sim
x^{-\lambda},$  ($\lambda$ varying in the range $0-0.5$). Their
results obey th following parameterization for the sea level muon
fluxes:
\begin{equation}
\phi_{\mu}(E_{\mu}) =
A\left(\frac{E_{\mu}}{1~GeV}\right)^{-(a+bx+cx^2+dx^3)}
\end{equation}
where $x= Log_{10}(E_{\mu}/GeV)$. The parameters are given in the
Table 1. We choose two representative sets corresponding to
$\lambda=0.1$ (GGV1) and $\lambda=0.5$ (GGV5).
\begin{table}[!ht]
\begin{tabular}{|l|l|l|l|l|l|l|l|r|}
\hline
Flux model &  $\lambda$ & A $\times 10^{-6}$& a & b & c$\times 10^{-2}$ & d$\times 10^{-3}$\\
\hline GGV1 &  0.1 & 3.12 & 2.70 & -0.095 & 1.49
& -0.2148 \\
GGV5 &  0.5 & 0.58 & 1.84 &
0.257 & -4.05 & 2.455 \\
\hline
\end{tabular}
\caption{GGV  parameters  for the prompt muon and antimuon fluxes.
 MRST PDFs are adopted and the scale is set to $\tilde{M}=2\tilde{\mu}=2 E_T = 2 (k_T^2 + m_c^2)$;
 $m_c=1.25$ GeV and  $k_T$ is the transverse momentum
of charm quark.} \label{t1}
\end{table}
As our last representative model, we consider flux calculation
within the saturation model proposed by Golec-Biernat and Wuthsoff
\cite{gbw,gbw1}. We follow \cite{mrs} in parameterizing the charm
production within this model. Within $10\%$ accuracy, we can adopt
the following parametrization
\begin{equation}
x\frac{d\sigma}{dx}{(p+~air\to c+..)} = Ax^{\beta}(1-x^{1.2})^n
\end{equation}
where $\beta = 0.05 - \ln(E/10~TeV)$ and
\begin{eqnarray}
n = \left\{\begin{array}{ll} 7.6 + 0.025\ln(E/10~TeV)&\textrm{$10^4<E < 10^8$ GeV}\\
7.6 + 0.012\ln(E/10~TeV)&\textrm{$10^8<E < 10^{11}$ GeV}
\end{array}\right.\end{eqnarray}

\begin{eqnarray}
A = \left\{\begin{array}{ll} 140 + (11\ln(E/0.1~TeV))^{1.65}&\textrm{$10^4<E < 10^8$ GeV}\\
4100 + 245\ln(E/10^5~TeV)&\textrm{$10^8<E < 10^{11}$ GeV}
\end{array}\right.\end{eqnarray}
A is expressed in nano-barns. For this model, we consider two cases:
GBW1 - where the protons are taken to be the primary and GBW2 -
where we include the effect of heavy elements also. On the basis of
arguments presented earlier, these two cases should have a different
nature, with the expectation that GBW2 should have a decreased
number of muons.

\par Figure 1 shows the results for muon fluxes in various models.
In Figure 2, we present the number of muons entering a typical
detector per solid angle in a span of five years. To have a quick
comparison, we present the same results in Table 2.
\begin{figure}[htp]
\hbox{\hspace{0cm} \hbox{\includegraphics[scale=0.65]{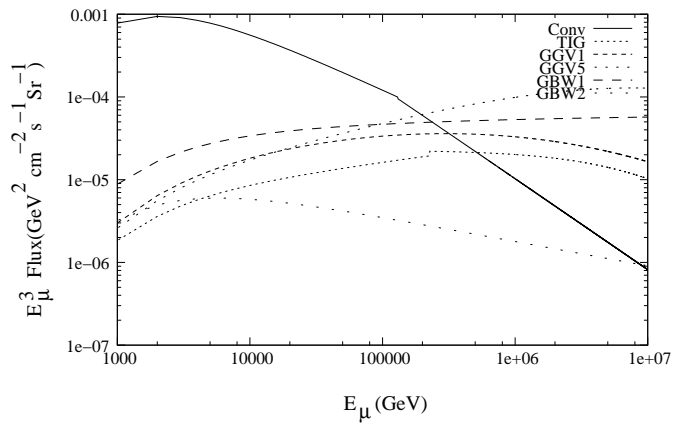}}}
\caption{${E_{\mu}}^{3} \times$  Flux vs. muon enegy $E_{\mu}$ in
different flux models entering the underground detector after
passing through the rock depth of $3.5 \times 10^5 gm/cm^2.$}
\label{fig1}
\end{figure}

\begin{figure}[htp]
\vskip 0.3cm \hbox{\hspace{0cm}
\hbox{\includegraphics[scale=0.65]{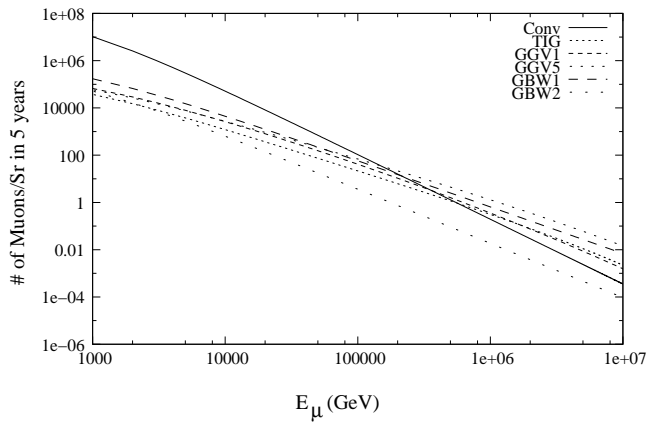}}}
\caption{Number of muons entering the 50 kT detector in 5 years per
solid angle vs. $E_{\mu}$ for different flux models. Energy loss in
the surrounding rock of depth $3.5 \times 10^5 gm/cm^2$ is taken
into account.} \label{fig2}
\end{figure}


\begin{table}[!ht]
\begin{tabular}{|l|l|l|l|l|l|l|l|r|}
\hline
$E_{\mu}$ & Conv & TIG & GGV1 & GGV5 & GBW1 & GBW2
\\ \hline
  1 & $10^7$ & 37461 & 66927& 58980 & 171625&
51564\\
  5 & 300322 & 3780 & 7920 & 7561 & 15160& 2770\\
  10 & 51282 & 1204 & 2618& 2700 & 4452 & 638\\
  50 & 696 & 74 &  154& 214& 219& 18 \\
  100 & 106 & 21 & 42 & 69 & 58&  4\\
  200 & 16 & 6 & 11 & 22 & 15  &  1\\
 300 & 5 & 3 & 5  & 11 &  7   & 0\\
  400 & 2 & 2 & 3  & 7 &  4   &  0 \\
  500 & 1 & 1 & 2 & 4 & 3  & 0 \\
  \hline
\end{tabular}
\caption{Number of muons per solid angle entering the detector in 5
years for various energies of the entering muon, $E_\mu$ (in TeV).}
\label{tab1}
\end{table}

{\bf Discussion and Conclusions~~} In Table 2 (and Figure 2), the
number of muons entering the detector are quoted as number per
steradian. However, a better quantity would be the total number of
muons. Very naively, for the down going muons that we are interested
in here, this would require multiplying these numbers by a factor
$2\pi$. However, in reality this would perhaps be overestimating the
number of muons. Therefore, a conservative multiplicative factor can
be $\# \pi$ where $1<\#<2$. From the figures and the table, it is
clear that direct measurement of muon spectrum using the pair meter
technique is capable of distinguishing between various models of
charm production. In particular, as expected, making the composition
heavier reduces the number of prompt muons (the curve GBW2 in Fig
2). The conventional contribution is dominant in this region.
However, due to the fact that the conventional flux is much better
known and the theoretical uncertainties are in better control there,
this should be not a serious trouble. As far as the prompt muons are
concerned, the conventional one should be viewed as a calculable
background. Further, from Table 2 we notice that $E_{\mu}\sim 200$
TeV seems as a practical cut-off within the set up assumed here.
However, let us remind ourselves that $E_{\mu}^{surface} \sim
5E_{\mu}$ and $E_{CR}\sim 20E_{\mu}^{surface}$ and the fact that
very high energy muons do not lose significant amount of energy
while traveling downwards. Therefore, a muon of given energy
entering the detector is actually probing the parent cosmic ray of
energy almost two orders of magnitude higher. We can thus expect to
probe the hadron dynamics in the region around the knee, and also
the possible change in composition, very easily with this technique.
Before concluding we would like to mention that there is an
indication of strangeness enhancement in high energy heavy ion
collisions from RHIC and CERN-SPS (for a survey of results, see
\cite{strangeness}). Taking clue from this, we expect the
contribution from the strange hadrons to increase. In particular,
for the heavy primary composition, the kaon component of the
conventional flux and prompt component from $D_s$ is expected to
change, though a detailed calculation is needed to quantify it.

\par In conclusion, we have explored the possibility of using pair
meter technique to measure the prompt muon spectrum. This can be
used to gain information about the charm production mechanism at
these energies and therefore yield invaluable information about
low-x behaviour of gluon distributions. The range probed here is
very different from what is expected to be probed at the LHC and
therefore this will provide an important set of data points in the
gluon distribution evolution plot. The results obtained are quite
encouraing and indicate that prompt muons (leptons in general) have
the capability of probing the charm (heavy hadron) production models
and also the hypothesis of composition change around the knee
region.


{\bf Acknowledgements~~}The work of N.M. is supported by
NSC95-2811-M-002-031, Taiwan (ROC) and the work of S.P. is supported
by Ministerio de Educacion y Ciencia, Spain. S.P. would like to
thank John Beacom for his comments on a previous paper of the author
which helped partially led to this work.


\end{document}